\documentclass[onecolumn, doublespace, 12pt fonts]{aastex6}

\newcommand{\beq}{\begin{equation}}
\newcommand{\eeq}{\end{equation}}

\shorttitle{Vertical Structure of Thin Disks}
\shortauthors{Gong \& Gu}

\begin{document}

\title{
Vertical Structure of Radiation-Pressure-Dominated Thin Disks: \\
Link between Vertical Advection and Convective Stability}

\author{Hong-Yu Gong\altaffilmark{1,2} and Wei-Min Gu\altaffilmark{1}}

\altaffiltext{1}{Department of Astronomy, Xiamen University,
Xiamen, Fujian 361005, China; guwm@xmu.edu.cn}
\altaffiltext{2}{Key Laboratory for the Structure and Evolution of
Celestial Objects, Chinese Academy of Sciences, Kunming 650011, China}

\begin{abstract}
In the classic picture of standard thin accretion disks, the viscous heating
is balanced by the radiative cooling through the diffusion process,
and the radiation-pressure-dominated inner disk suffers convective instability.
However, recent simulations have shown that the vertical advection process
owing to the magnetic buoyancy can make significant contribution to
the energy transport.
In addition, no convective instability has been found by comparing the
simulation results with the local convective stability criterion. 
In this work, following the spirit of simulations, we revisit the vertical
structure of radiation-pressure-dominated thin disks
by including the vertical advection process.
Our study indicates a link between the additional energy transport
and the convectively stable property.
Thus, the vertical advection not only has significant contribution to
the energy transport, but also plays an important role to make the disk
convectively stable.
Our analyses may be helpful to understand the discrepancy between
the classic theory and simulations on standard thin disks.
\end{abstract}


\section{Introduction}

The standard thin accretion disk model under the alpha description
was constructed by Shakura and Sunyaev \citep{SS73}, which has been
successfully applied to X-ray binaries and active galactic nuclei
\citep[for a review, see][]{Frank02,Kato08}.
However, when we compare the theoretical results with observations
and simulations, there exist some basic problems
such as thermal stability and convective stability.
Recent simulations \citep{Hirose09,Jiang13} on thin disks based on
a shearing box showed that,
for high mass accretion rates around $0.1 \dot M_{\rm Edd}$, where
$\dot M_{\rm Edd}$ is the Eddington accretion rate, the inner disk
is radiation pressure dominated, and the vertical advection can be
an efficient process for energy transport, which is probably
related to the magnetic buoyancy \citep{Hirose09,Jiang13}.
In addition, the local convective stability criterion,
$dS/dz > 0$, where
$S$ is the entropy, is well satisfied according to the simulation results.
On the other hand, in the classic theory of thin disks, the vertical
energy transport is completely dominated by the diffusion process.
Moreover, when the radiation pressure dominates over the gas pressure,
the local convective stability criterion $dS/dz > 0$ is not satisfied
and therefore the disk may suffer convective instability \citep{Sadow11}.
Then a question arises that, is there any link between the vertical
advection and the convective stability?

In the present work, we will revisit the vertical structure and energy
transport of radiation-pressure-dominated thin disks by including
the possible vertical advection process. In addition,
based on the simulation results, we will take into account the
local convective stability criterion to modify the set of equations.
The rest part is organized as follows. Equations and boundary conditions
are described in Section~2. Numerical results are shown in Section~3.
Conclusions and discussion are made in Section~4.

\section{Equations and boundary conditions}

The set of equations for the vertical structure of thin disks
is based on the alpha-stress assumption. The gas pressure $P_{\rm gas}$
and the radiation pressure $P_{\rm rad}$ and their derivates
take the following expressions:
\beq
P_{\rm gas} = \frac{\rho k_{\rm B} T}{\mu m_{\rm p}} \ ,
\eeq
\beq
P_{\rm rad} = \frac{1}{3}aT^4 \ ,
\eeq
\beq
\frac{d P_{\rm rad}}{dz} = - \frac{\kappa_{\rm es} \rho F_z}{c} \ ,
\eeq
\beq
\frac{d P_{\rm gas}}{dz} =  -\rho\Omega_{\rm K}^2 z
+ \frac{\kappa_{\rm es} \rho F_z}{c} \ ,
\eeq
where $\rho$ is the mass density, $T$ is the temperature, $k_{\rm B}$ is
the Boltzmann constant, $m_{\rm p}$ is the mass of a proton.
The Keplerian angular velocity $\Omega_{\rm K}$ is written as
$\Omega_{\rm K} = \sqrt{GM_{\rm BH}/R(R-R_{\rm g})^2}$ under the
well-known Paczy\'{n}ski-Wiita potential
$\Psi_{\rm PW} = - GM_{\rm BH}/(R - R_{\rm g})$ \citep{PW80},
where the gravitational radius is defined as
$R_{\rm g} \equiv 2GM_{\rm BH}/{c^2}$.

In the standard disk model, the angular velocity is assumed to be
Keplerian, i.e., $\Omega = \Omega_{\rm K}$.
Moreover, the local energy balance at each radius
is that the viscous heating rate equals the radiative cooling rate,
and the vertical energy transport
is dominated by the photon diffusion process.
Here, following the spirit of simulation results,
the vertical advection may have an additional contribution to
the energy transport. Thus,
the vertical flux $F_z$ due to the diffusion is expressed as
\beq
\frac{dF_z}{dz} = - \tau_{r\phi} \Omega_{\rm K} g_{\ast}
- \frac{dF_{\rm adv}}{dz} \ ,
\eeq
where $F_{\rm adv}$ is the vertical flux owing to the vertical
advection process, and the parameter $g_{\ast}$ takes the form
$g_{\ast} \equiv -d\ln \Omega_{\rm K}/d\ln R = 3/2 + 1/(R/R_{\rm g} -1)$.
The $r\phi$-component of the shear stress $\tau_{r\phi}$ is assumed to be
proportional to the total pressure, i.e.,
$\tau_{r\phi} = - \alpha (P_{\rm gas} +P_{\rm rad})$, where
$\alpha$ is a constant parameter.

The total entropy $S$ is the sum of gas and radiation, which is written as
\[
S = S_{\rm gas} + S_{\rm rad}
= \frac{3}{2}\frac{k_{\rm B}}{\mu m_{\rm p}}
\ln \left( \frac{P_{\rm gas}}{\rho^{5/3}} \right)
+ \frac{4}{3}\frac{aT^3}{\rho} + {\rm const.}
\]
It is well-known that the local convective stability criterion
can be expressed as
\[
\frac{dS}{dz} > 0 \qquad (z>0) \ ,
\]
which is equivalent to the relationship between the radiative gradient
$\nabla_{\rm rad}$ and the adiabatic gradient $\nabla_{\rm ad}$:
\[
\nabla_{\rm rad} < \nabla_{\rm ad} \ ,
\]
where

\[
\nabla_{\rm rad} \equiv
\left( \frac{\partial\ln T}{\partial\ln P} \right)_{\rm rad} \ , \qquad
\nabla_{\rm ad} \equiv
\left( \frac{\partial\ln T}{\partial\ln P} \right)_S \ .
\]
With the above expression of the total entropy $S$, we can derive the
explicit form of $\nabla_{\rm ad}$ as
\[
\nabla_{\rm ad} =
\frac{4-3\beta}{1.5\beta^2 + 12\beta(1-\beta) + (4-3\beta)^2} \ ,
\]
where $\beta$ is defined as
$\beta \equiv P_{\rm gas}/(P_{\rm gas}+P_{\rm rad})$.

As mentioned in the first section, simulations have not shown convective
instability, which indicates that the relationship
$\nabla_{\rm rad} \leqslant \nabla_{\rm ad}$ may be well satisfied
in the disk.
Following this spirit, we assume the thermodynamical
gradient to be
\beq
\frac{d\ln T}{d\ln P} = \min (\nabla_{\rm rad},\lambda \nabla_{\rm ad}),
\eeq
where $\lambda \leqslant 1$ can guarantee the disk to be convectively
stable.

The system consists of six equations, Equations~(1-6), for the six unknown
variables $\rho$, $T$, $P_{\rm gas}$, $P_{\rm rad}$, $F_z$, and $F_{\rm adv}$.
There are four first-order differential equations in this system.
In addition, the position of photosphere $H_{\rm phot}$ is unknown.
Thus, totally five boundary conditions are required to
solve the system between the equatorial plane ($z=0$) and
the photosphere ($z=H_{\rm phot}$).

On the equatorial plane there exist two natural boundary conditions:
\beq
F_z = 0 \ ,
\eeq
\beq
F_{\rm adv} = 0 \ .
\eeq
At the photosphere, the other three boundary conditions can be derived:
\beq
- \kappa_{\rm es} \rho^2 \left( \frac{d\rho}{dz} \right)^{-1} = 1 \ ,
\eeq
\beq
F_z = 2 \sigma T^4 \ ,
\eeq
\beq
F_z + F_{\rm adv} = \frac{1}{4\pi} \dot M \Omega_{\rm K}^2 f_{\ast} g_{\ast} \ ,
\eeq
where Equation~(9) can be regarded as the definition of the photosphere position.
The five boundary conditions (7-11) together
with Equations (1-6) enable us to solve the system and derive the
vertical structure.

\section{Numerical results}

\begin{figure}
\centering
\plotone{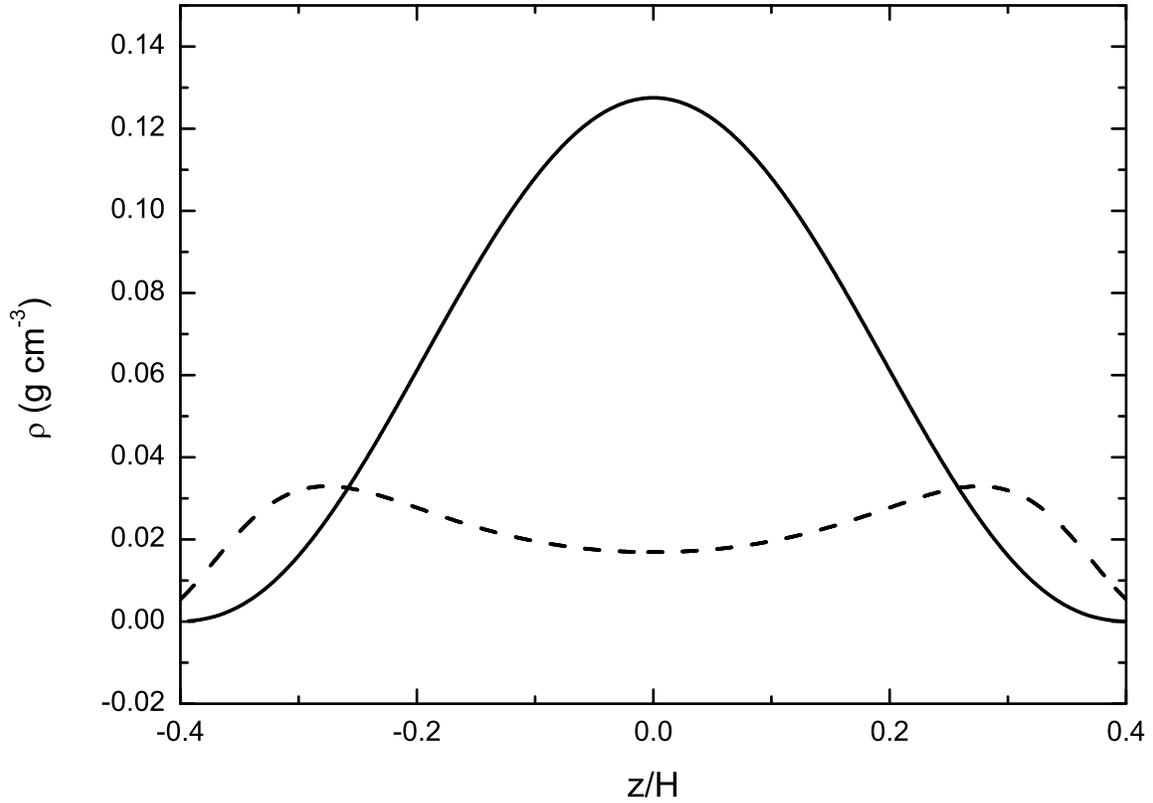}
\caption{
Vertical density profiles for the model including vertical advection
(solid line) and the standard thin disk model (dashed line).
}
\end{figure}

Following \citet{Hirose09}, we define a vertical thickness $H$ as
\[
H=\frac{3}{8\pi}\frac{\kappa_{\rm es} \dot M}{c} \ ,
\]
which is used as a scale of the height.
For simplicity, we choose $\lambda = 1$ in Equation~(6) for our numerical
calculations. The other parameters are
$M_{\rm BH} = 10 M_{\sun}$, $\dot M = 0.1 \dot M_{\rm Edd}$,
and $\alpha = 0.01$.

In order to directly compare the structure including vertical
advection process with that in the classic picture, we made numerical
calculations for both of these two models.
Figure~1 shows the vertical density profile, where the solid line corresponds
to the results including vertical advection, and the dashed line corresponds
to the results of the classic model. 
The solid line has a peak at $z=0$ whereas the dashed line has two peaks at
$z/H \approx \pm 0.3$. The peculiar shape of the dashed line, i.e.,
increasing $\rho$ with $z$ in the range $0 < z/H < 0.3$, indicates that
the disk suffers convective instability.
The well-known local convective stability criterion,
$dS/dz>0$ for $z>0$ (or $dS/dz<0$ for $z<0$) may work well in geometrically
thin disks, where the radial velocity is low and therefore the advection
effects may be negligible.
Even though the entropy profile is not plotted in Figure~1, it is obvious
that the entropy $S$ will decrease with increasing $z$ in the range
$0 < z/H < 0.3$ due to the peculiar profile of density.
Thus, the local convective stability criterion is not satisfied
and therefore the disk is likely to be convectively unstable,
as previously investigated by \citet{Sadow11}.
On the contrary, the profile of the density in the case including vertical
advection (solid line) is similar to that of simulations \citep{Hirose09},
which shows a continuously decreasing density with increasing $z$ for $z>0$.
Thus, a convectively stable disk is quite possible.

\begin{figure}
\centering
\plotone{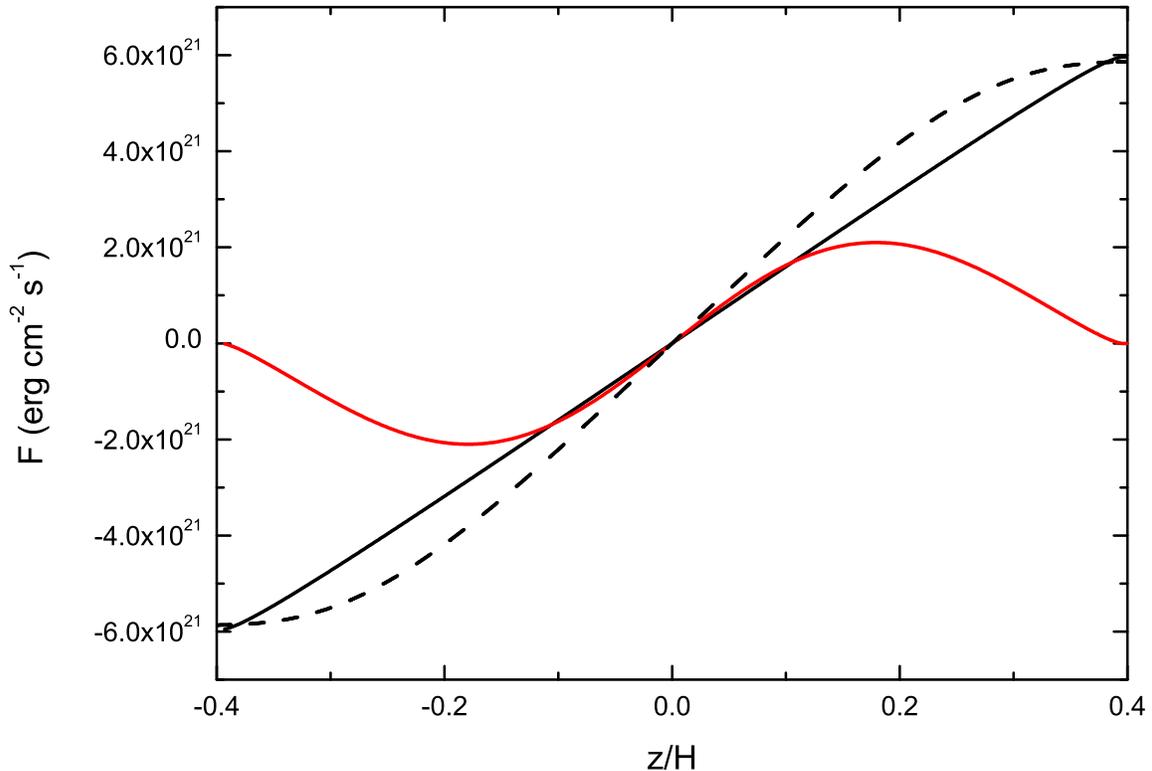}
\caption{
Vertical profiles of the diffusive flux $F_z$ (black solid line)
and the advective flux $F_{\rm adv}$ (red solid line) for the model
including vertical advection, and the profile of $F_z$
for the standard thin disk model (black dashed line).
}
\end{figure}

Figure~2 provides the vertical profiles of flux, where the black
dashed line corresponds to the diffusive flux $F_z$ in the classic thin
disk model, and the two solid lines correspond to the case including the
vertical advection, where the black and red lines show the variations of
the diffusive flux $F_z$ and the advection flux $F_{\rm adv}$, respectively.
It is seen that the vertical advection has significant
contribution to the total radiation flux, in particular for the region
near the equatorial plane ($0 < z/H < 0.1$). In some regions on the 
right part ($z>0$), the red line is even higher than the black solid line,
which means that $F_{\rm adv}$ can dominate over $F_z$.
In addition, the profile of $F_{\rm adv}$ in Figure~2 is quite similar to
that in simulations \citep[e.g.,][]{Hirose09}.
Since our numerical calculation is based on a convectively stable disk,
the results may indicate a link between the energy transport due to
vertical advection and the convectively stable property.
The physical reason for the link is probably that, the additional
energy transport decreases the entropy in the region near the equatorial
plane and therefore the entropy can keep to increase with increasing
vertical height, which satisfies the local convective stability
criterion.

\begin{figure}
\centering
\plotone{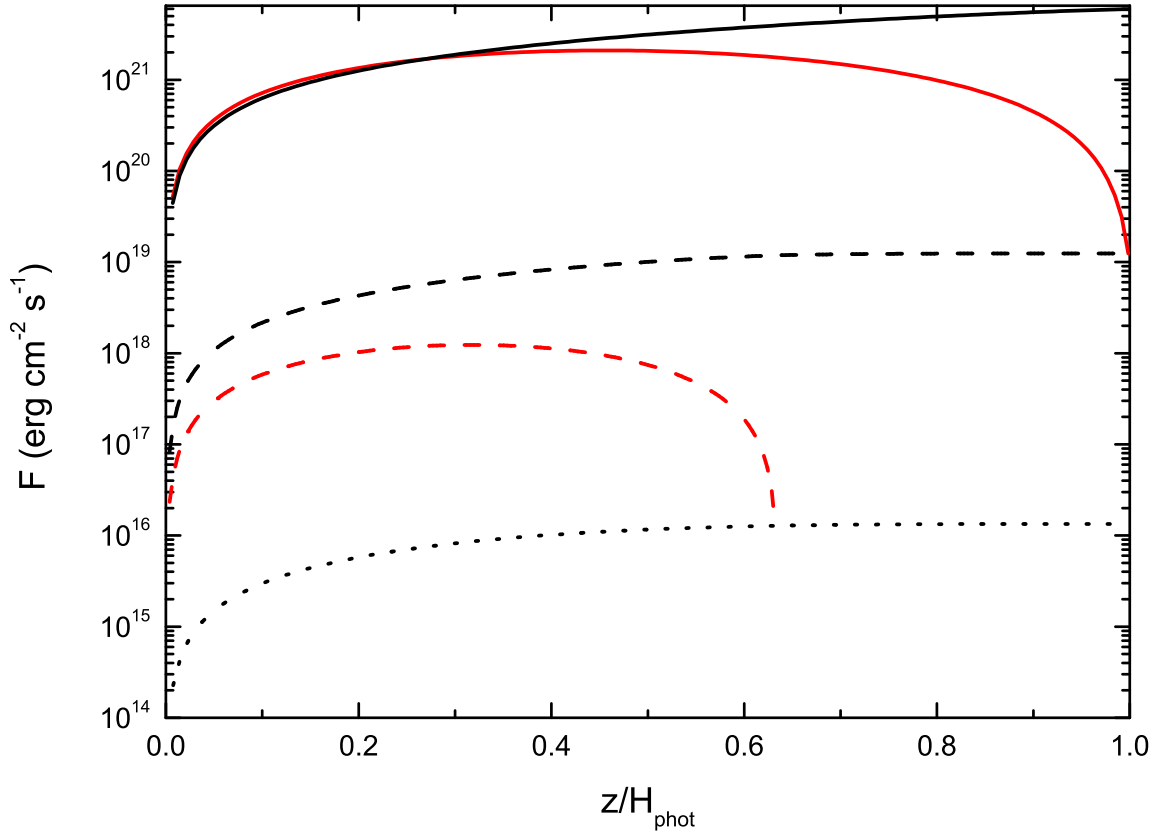}
\caption{
Vertical profiles of the diffusive flux $F_z$ (black lines) and the advective
flux $F_{\rm adv}$ (red lines) for $R = 10 R_{\rm g}$ (solid),
$100 R_{\rm g}$ (dashed), and $1000 R_{\rm g}$ (dotted).
}
\end{figure}

Taking $\dot M = 0.1 \dot M_{\rm Edd}$ as a typical accretion rate, we also
investigate the strength of vertical advection for different radii.
Obviously, the ratio of the gas pressure to the total pressure $\beta$
will increase with increasing radius for a fixed $\dot M$.
In the outer part where gas pressure is dominant, the vertical advection
may be negligible. In other words, the diffusion process is the dominant
energy transport and the disk is well convectively stable.
However, in the inner part where the radiation pressure dominates over the gas
pressure, the diffusion and the vertical advection may both be of
importance for the energy transport.

Figure~3 shows the profiles of diffusive flux (black) and advective flux (red)
at the three radii $R = 10 R_{\rm g}$ (solid lines), $100 R_{\rm g}$
(dashed lines), and $1000 R_{\rm g}$ (dotted line).
Here, we take the height of photosphere $H_{\rm phot}$ as the length unit.
The red solid line shows that, at $R = 10 R_{\rm g}$, the advective flux
$F_{\rm adv}$ covers the range from the equatorial plane ($z=0$) to
the photosphere ($z=H_{\rm phot}$).
At a larger radius $R = 100 R_{\rm g}$, the red dashed line shows
a smaller vertical range for $F_{\rm adv}$ ($\la 2 H_{\rm phot}/3$). 
Moreover, for a sufficiently large radius $R = 1000 R_{\rm g}$, Figure~3 shows
that the advective flux disappears and the diffusion is the only
mechanism for the vertical energy transport, as shown by the dotted line.
Thus, for thin disks, the two conditions for the occurrence of vertical
advection are high accretion rates $\dot M \ga 0.1 M_{\sun}~{\rm s}^{-1}$
and small radii $R \la 100 R_{\rm g}$.

\section{Conclusions and discussion}

In this work, we have revisited the vertical structure of
radiation-pressure-dominated thin disks by taking into account the role of
vertical advection process. Our study has shown that
the vertical advection not only has significant contribution to the
energy transport, but also plays an important role to make the disk
convectively stable.
Thus, a link may exist between the vertical advective energy transport
and the convectively stable property.
The physical reason for the link is probably that, the additional
energy transport decreases the entropy in the region near the equatorial
plane and therefore the entropy can keep to increase with increasing
vertical height, which satisfies the local convective stability
criterion. 
Our work may be helpful to understand the discrepancy between the classic
theory and simulation results.

We would point out that, the detailed study of convective stability
may require global rather than local stability analyses. For example,
\citet{Abram93} demonstrated that when the viscosity is taken fully
into account, stability analyses cannot be discussed within the
framework of a local analysis, and a fully global treatment is required.
Moreover, the global stability analyses of vertical convection of a thin
gaseous disk were performed by several works \citep[e.g.,][]{Ruden88}.
On the other hand, it is known that the advection may play an important
role in stabilizing the disk against dynamic, thermal, and viscous
perturbations. For instance, the geometrically thick disk without
radial motion \citep{Abram80,PW80} may suffer the Papaloizou-Pringle
instability \citep{PP84}, which is a dynamic instability based on the
acoustic perturbations propagating between two boundaries in a differential
rotating system. Later, by
including the advection terms, \citet{Blaes87} found that all the
unstable modes for the purely rotating flow are quickly stabilized
by the advection process. Furthermore, \citet{Abram88} proposed the
well-known slim disk model (or named as the optically thick,
advection-dominated accretion disk) for super-Eddington accretion systems.
The slim disk was found to be dynamically stable, thermally stable,
and viscously stable, which may be related to the radial advection process.
In the present work, we focus on the stability of geometrically thin disks,
where the radial velocity ($v_R \approx - \alpha (H/R)^2 v_{\rm K}$,
where $v_{\rm K}$ is the Keplerian velocity) is quite low, and therefore
the effects of advection may also be quite weak.

In a previous work, \citet{Gu12} showed that, for high mass
accretion rates $\dot M \ga \dot M_{\rm Edd}$ where the radiation
pressure completely dominates over the gas pressure, the disk
is likely to be convectively stable without including the energy
transport through the vertical advection.
The physics of the stable property is probably related to the radial
advection effects and the geometrically thick structure.
On the other hand, in recent years many global simulation works have been done
on the super-Eddington accretion flows
\citep[e.g.,][]{Ohsuga05,Jiang14,Sadow15}.
The simulations of \citet{Jiang14} revealed the importance of
the vertical advection, which can essentially enhance the radiative
efficiency. In addition, outflows may play another important role in such
flows \citep{Jiang14,Sadow15}.
In our opinion, the theory of super-Eddington accretion flows is worth
further investigation.

\acknowledgments

The authors would thank Yan-Fei Jiang for providing the vertical profile
of entropy in simulations, and thank the referee for helpful comments
that improved the paper.
This work was supported by the National Basic Research Program of China
(973 Program) under grants 2014CB845800,
the National Natural Science Foundation of China under grants 11573023,
11333004, and 11222328,
and the CAS Open Research Program of Key Laboratory for the Structure
and Evolution of Celestial Objects under grant OP201503.

\end{document}